\documentclass[letterpaper]{article} 
\usepackage{amsmath}
\usepackage{amssymb}
\usepackage{algorithm}
\usepackage{algorithmic}
\usepackage{booktabs}
\usepackage{array}
\usepackage{aaai24}  
\usepackage{times}  
\usepackage{helvet}  
\usepackage{courier}  
\usepackage[hyphens]{url}  
\usepackage{graphicx} 
\usepackage{natbib}  
\usepackage{caption} 
\usepackage{algorithm}
\usepackage{algorithmic}

\usepackage{newfloat}
\usepackage{listings}
\usepackage{bibentry}
\DeclareCaptionStyle{ruled}{labelfont=normalfont,labelsep=colon,strut=off} 
\floatstyle{ruled}
\newfloat{listing}{tb}{lst}{}
\floatname{listing}{Listing}
\usepackage[utf8]{inputenc}
\usepackage{newunicodechar}
\DeclareUnicodeCharacter{FF0C}{\text{，}}

\title{An Unsupervised Domain Adaptation Method for Locating Manipulated Region in Partially Fake Audio}
\author {
    Siding Zeng\textsuperscript{\rm 1, \rm 2},
    Jiangyan Yi\textsuperscript{\rm 2},
    Jianhua Tao\textsuperscript{\rm 3},
    Shan Liang\textsuperscript{\rm 2},
    Yujie Chen\textsuperscript{\rm 2},
    Chenglong Wang\textsuperscript{\rm 2},
    Yong Ren\textsuperscript{\rm 2},
    Xiaohui Zhang\textsuperscript{\rm 2}
}
\affiliations {
    \textsuperscript{\rm 1}University of Chinese Academy of Sciences\\
    \textsuperscript{\rm 2}Institute of Automation, Chinese Academy of Sciences\\
    \textsuperscript{\rm 3}Department of Automation, Tsinghua University\\
    zengsiding2023@ia.ac.cn, jhtao@tsinghua.edu.cn, jiangyan.yi@nlpr.ia.ac.cn
}

\begin{document}

\maketitle

\begin{abstract}
 When the task of locating manipulation regions in partially fake audio (PFA) involves cross-domain datasets, the performance of deep learning models drops significantly due to the shift between the source and target domains. To address this issue, existing approaches often employ data augmentation before training. However, they overlook the characteristics in the target domain that are absent in the source domain. Inspired by the mixture-of-experts model, we propose an unsupervised method named Samples mining with Diversity and Entropy (SDE). Our method first learns from a collection of diverse experts that achieve great performance from different perspectives in the source domain, but with ambiguity on target samples. We leverage these diverse experts to select the most informative samples by calculating their entropy. Furthermore, we introduce a label generation method tailored for these selected samples that are incorporated in the training process in the source domain integrating the target domain information. We applied our method to a cross-domain partially fake Audio Detection dataset, ADD2023Track2. By introducing 10\% of samples from the target domain, we achieved an F1 score of 43.84\%, which represents a relative increase of 77.2\% compared to the second-best method.
\end{abstract}

\section{Introduction}
AI generated content technology has witnessed swift progress in recent years, particularly in speech-related applications such as text-to-speech (TTS)  \citep{wang2023neural,huang2022prodiff,tan2024naturalspeech} and voice conversion (VC)\citep{chan2022speechsplit2,patel2015combining,chen2021again}.
Although these technologies have brought about convenience, they have also brought about the forgery and deception of voices, which pose significant security threats to social security, privacy security, and property security \citep{lorenzo2018voice,kinnunen2018spoofing,lorenzo2018can}. 
As an important branch of voice forgery and deception, partially fake audio (PFA) \cite{yi2023add} has been receiving an increasing focus in recent years. Attackers can contaminate original voice segments with synthesized or authentic vocal utterances. Such low-cost modifications can easily alter the semantics of sentences.
For instance, attackers can easily modify a single word such as time, place, and characters in sentence to dramatically change the semantics. Furthermore, if attackers have knowledge of phonology, they can manipulate vowels and even consonants such as “pan”, “pin”, “pen”,in English or “sha”, “la”, “pa”, in Chinese, which are smaller than the word level. These PFAs contain a substantial amount of genuine information, making them undetectable by models designed for authenticating entire segments of speech.
\begin{figure}[t]
    \centering
    \includegraphics[width=\linewidth]{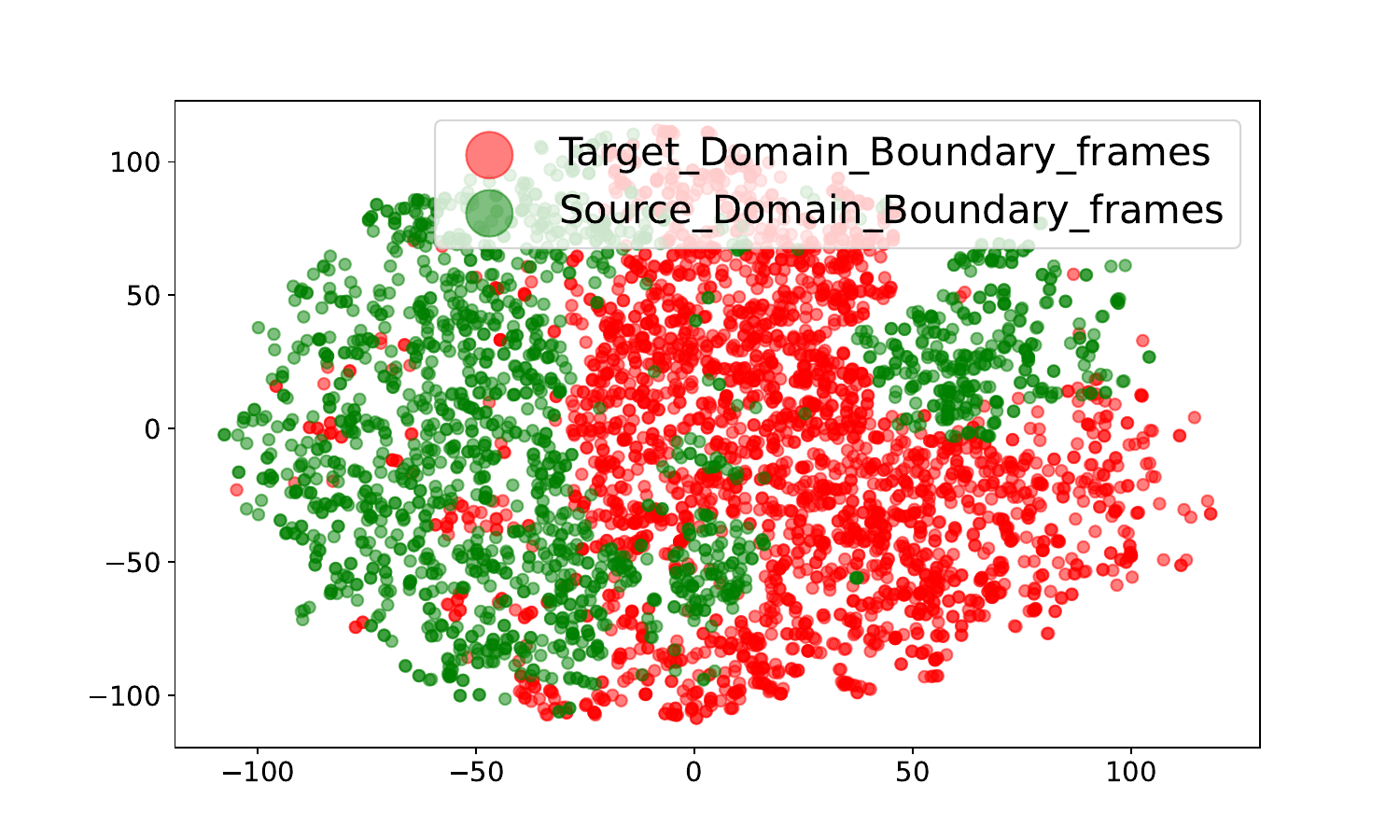}
    \label{fig:enter-label}
    \caption{The visualization of the feature distribution of HAD training dataset(source domain) and ADD2023 test dataset(target domain)}
\end{figure}
However, compared to PFA detection, which tries to identify the authenticity of voices containing a substantial amount of genuine content, locating the manipulated region of PFA is more meaningful. PFA detection focuses on sentence-level positive detection, determining whether a sentence has been manipulated, while locating the manipulated region of PFA emphasizes frame-level positive detection, assessing whether a vowel or even a single frame has been manipulated, which can be more challenging. Therefore, defending against such fine-grained partially fake scenarios poses significant challenges for defenders. 
Existing research on the localization of manipulated regions has shown satisfactory performance on certain in-domain partially fake audio datasets.\cite{9746596,xie2024efficient}. Nevertheless, their efficacy markedly declines when applied to cross-domain partially fake audio datasets. This phenomenon, known as out-of-distribution, arises when there is a shift between the source domain and the target domain. 
In addressing this issue, researchers have proposed many methods in the field of anti-spoof tasks for fake audio. However, domain adaptation for PFA has not yet received widespread attention. When participants in the ADD2023 competition face cross-domain challenges, they often resort to data augmentation as a coping strategy\cite{martin2023vicomtech,li2023multi,cai2024integrating}, such as adding noise or processing segments through voice conversion models.

However, methods based on data augmentation still only consider information from the source domain and do not assess the information missing in the target domain.
The urgent problem is that, under the standard empirical risk minimization (ERM) paradigm, machine learning algorithms tend to choose solutions based on the most salient features without considering alternatives. Learning from one distribution and applying it to another can cause model failure. Any method based on the source domain struggles to be effective when there is a significant distribution shift between the source and target domains, because machine learning algorithms have never encountered characteristic from the target domain.
Therefore, it is necessary to incorporate samples from the target domain for domain adaptation.
In such cases, selecting the highest quality samples becomes crucial. Many methods proceed through various analyses and then introduce samples of the target domain. However, they overlook the relativity of information; information already present in the source domain should not be the focus of learning, while information absent in the source domain should be emphasized.
However, in the task of PFA, the situation is somewhat different. The normal domain adaptation task generally refers to adaptation between two domains. However, as for the partially fake audio dataset, it is much more difficult. Due to the recording environment, devices, compression methods, and other factors, there will exhibit distribution differences among the unprocessed utterances. Moreover, considering the variety of methods for generating manipulated fake audio, coupled with the manipulation using non-synthetic real audio, the distribution of audio sources in "partially fake audio" becomes even more complex. This complexity makes the task of "Locating Manipulated Region in partially fake audio" a more challenging cross-domain issue.

So we propose our method for the task of Locating Manipulated Region in partially fake audio (PFA) in cross-domain scenarios, structured in three phases. In the first phase, we employ a reverse knowledge distillation approach to learn a collection of diverse experts, which is inspired by the Mixture of Experts model. Although these experts share the same structure, our reverse distillation process constraints allow them to achieve high performance on source domain but interpret the source domain from multiple perspectives. Then, with the assistance of these diverse experts, we aim to find samples that contain the subset of the symmetric difference between the collection of information in the source domain by calculating the entropy of samples relative to diverse experts, which can be viewed as characteristics present in the target domain but absent in the source domain. Finally, we transfer information from the target domain by swapping voice active segments and generating labels, allowing these samples to participate in the training process.
By incorporating 10\% the most informative samples from the target domain into our analysis, we attained an F1 score of 43.84\%. This score signifies a 77.2\% relative improvement over the performance of the second-best method.
Furthermore, the results of the Ablation Experiment demonstrate the unique contribution of each step of SDF, indirectly validating the effectiveness of our method.

We make contribution as follows:
\begin{itemize}
\item
we propose a method training for diverse experts based on reverse knowledge distillation.
\end{itemize}
\begin{itemize}
\item
We propose a method that introduces information entropy to evaluate the most informative samples based on a set of diverse experts.
\end{itemize}
\begin{itemize}
\item
We proposed a soft-label generation method tailored for the most informative samples in PFA tasks.
\end{itemize}

The remainder of this survey is structured as follows. In Section 2, we introduced the relevant work and progress of the PFA. In Section 3, we first defined some notations, and then proposed our method. Experiments, results, and discussion are reported in Section 4. Finally, we concluded the paper in Section 5.

\section{related work}
since research on the detection of synthetic speech has already become very sophisticated\cite{zhang2023you,zhang2024remember,zhang2023multimodal,zhao2022emofake,wang2023low,zhang2023adaptive,wang2024multi,chen2024rawbmamba}, numerous methods and datasets have been developed for partially fake audio tasks .  Yi et al. \cite{yi2021half, khanjani2023audio}constructed the first partially spoofed audio dataset named HAD, which replaces some nature segments with synthesized speech segments having different semantic content. The Audio Deep Synthesis Detection (ADD) challenge 2022 \cite{yi2022add} involves a detection track containing partially spoofed audio. The organizers provided only real and entirely fake data for training, and the task was target to detect fake at utterance level. In this challenge, notable detection performance was achieved by fine-tuning pretrained self-supervised learning (SSL) models with different back-end classifiers \cite{lv2022fake,liu2022deep}. Boundary detection task was only introduced as a proxy task to achieve utterancelevel detection \cite{cai2023waveform,wu2022partially}.
 
Simultaneously, Zhang et al.\cite{zhang2021initial} construct a speech database called ‘Partial Spoof’ designed for Partially scenarios. The above two datasets are the beginning of the research for PF scenario in ADD task. Afterward, Zhang et al. \cite{zhang2021multi} propose the SELCNN network to enhance the ability of the accuracy of the utterance. Lv et al. \cite{lv2022fake} use Wav2Vec2 (W2V2) \cite{baevski2020wav2vec} as front-end, ECAPA-TDNN \cite{desplanques2020ecapa} as back-end achieving the first rank in ADD 2022 Track 2\cite{yi2022add}. Although the above research shows effectiveness at the utterance level detection in PF, they do not pinpoint specific segments with precision. Thus, Zhang et al. \cite{zhang2022partialspoof} extended the previous utterance-level PF dataset labels to frame-level and proposed corresponding W2V2-based countermeasures to enhance frame-level detection capability.and scholars from Inner Mongolia University have proposed a method to detect high-frequency features at splicing sites using wavelet functions\cite{9746596}.
 Moreover, the ADD 2023 \cite{yi2023add} introduced a PSAL track, which has resulted in numerous promising works \cite{martin2023vicomtech,li2023multi,cai2024integrating}. Among them, an effective approach was proposed to achieve the best localization score by integrating the decisions from three countermeasure (CM) systems including boundary detection, frame-level fake detection, and utterance-level fake detection \cite{cai2024integrating}
 Recently, an approach incorporating a contrastive learning module and temporal convolution was proposed to effectively capture better features for localization \cite{xie2024efficient}.
 \begin{figure*}[ht]

  \centering
  \includegraphics[width=\textwidth]{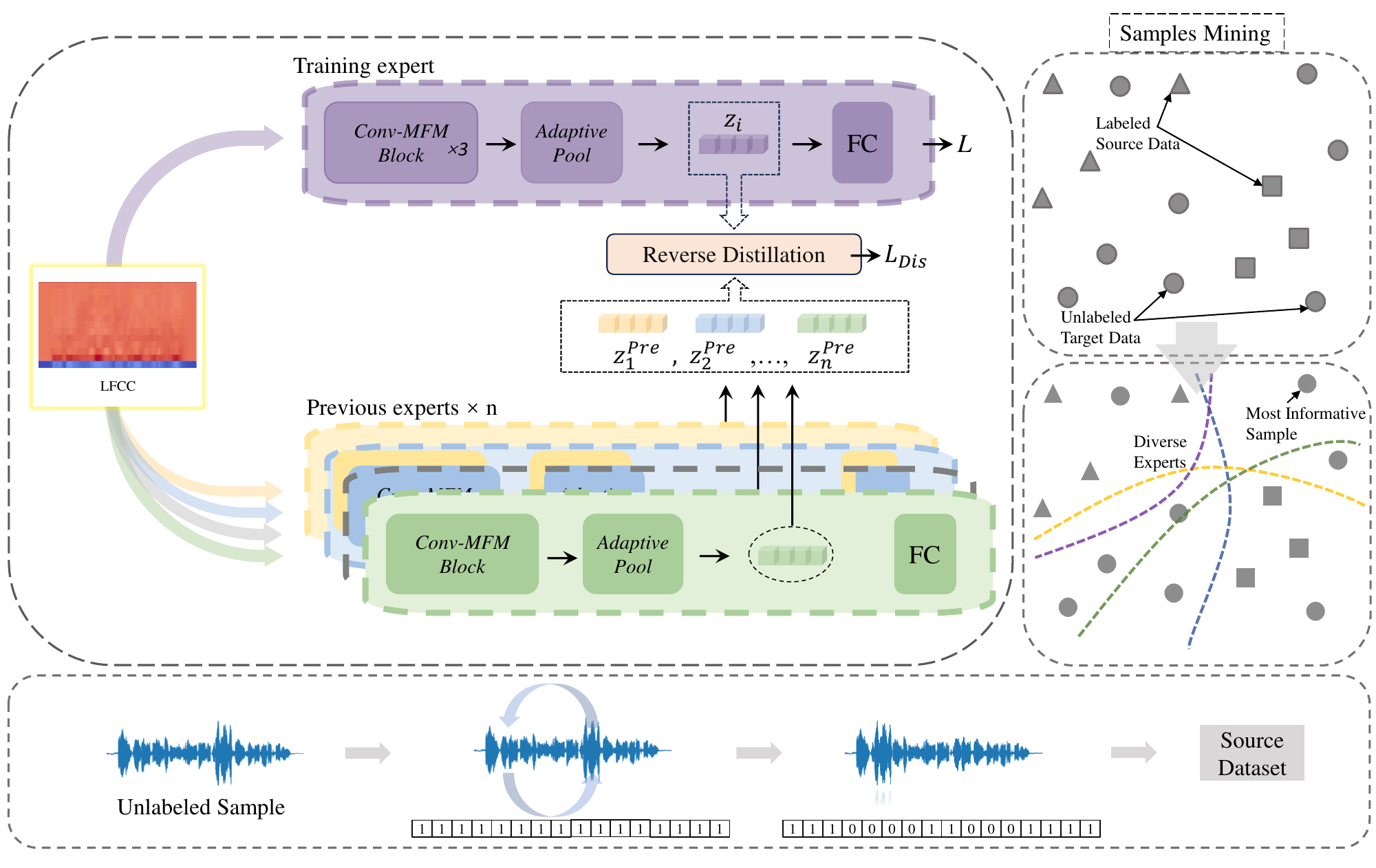} 
  
  \label{fig:bigfigure}
  \caption{The workflow of SDF: On the left is our process of training diverse experts through reverse knowledge distillation. On the right is the process of mining the most informative samples in the target domain using diverse experts. Below is our process for generating labels for these most informative samples, which ultimately participate in the training.}
\end{figure*}
\section{proposed method}
\subsection{Model Always with Ambiguity}
Consider a specifically situation, we train a model \(f\) that takes input feature of every single frame as input \(x \in \mathcal{X}\) and predicts its corresponding label \(y \in \mathcal{Y}\). We train \(f\) using a labeled dataset \(DS = \{(x_1, y_1), (x_2, y_2), \ldots\}\) drawn from the data distribution \(p_S(x, y)\). The model $f$ learns to adjust its parameters by approximately minimizing the predictive risk $E_{p_S}(x, y) \left[ \lVert f(x) - y \rVert \right]$ on the source data distribution $p_S(x, y)$, aiming to reduce its prediction error for the training dataset labels, thereby achieving the required performance on the current data distribution. The collection of parameters of the model that learns under such conditions above is defined as $\mathcal{F}$, which is also considered as the model's set of interpretations for the source domain.Then each \(f\in \mathcal{F}\) is evaluated based on its predictive risk on held-out samples from \(p_S(x, y)\). 

However, this approach can be less than ideal, leading to model failures when there is a shift in the data distribution, The standard paradigm of empirical risk minimization performs poorly on cross-domain tasks, as it tends to select solutions based on the most salient features (i.e., directions in the parameter space where the loss decreases most rapidly) without considering other possibilities. This simplicity bias occurs even when training an ensemble, as machine learning models still tends to choose the easy way to learn. This lead to that methods concentrate exclusively on one solution, ignoring the possibility of multiple interpretations. Since the prominent features in the source domain are not suitable for the target domain, such decisions frequently lead to failure. Although many recent methods have improved robustness in distribution shift settings, standard procedures such as regularization and cross-validation , encourage this generalization, they fail in the presence of severe under-specification.

 In many distribution shift scenarios \cite{koh2021wilds}, the overall data distribution can be modeled as a mixture of domains, where each domain \(d \in \mathcal{D}\) corresponds to a fixed data distribution \(p_d(x, y)\). Even if a model \(f\) generalizes to unseen data sampled from the same distribution \(p_S(x, y)\), performance often deteriorates under distribution shift conditions, when evaluated on target data sampled from a different distribution \(p_T(x, y)\).Datasets frequently lacking specific details, implying that several reasonable assumptions could explain the known data distribution just as effectively, with the available information not providing adequate evidence to favor one assumption over the others. This uncertainty results in machine learning models arbitrarily choosing one potential explanation for the presented dataset within the parameter space, devoid of any external guidance. These choices might not be ideal, leading to the models' underperformance upon shifts in data distribution, a situation often encountered in practical applications.
This occurs because datasets often lack detailed information, allowing for multiple reasonable interpretations that can explain the data equally well. In such instances, the data does not offer extra evidence to select the correct interpretation over others, thereby failing to resolve this ambiguity.

\subsection{Comprehensive Coverage}
We have specific expectations regarding the model's fit to the dataset. Under the condition that this anticipated performance is met, all combinations of parameters adopted by the model in the parameter space constitutes the model's complete interpretation of the dataset.In this discussion, we define a critical concept within the supervised learning framework: the best model Set: Let \(p(x, y)\) be a distribution defined over \(X \times Y\), and \(\mathcal{F}\) be the set of models achieving the expectation under \(\epsilon\), where the \(\epsilon\)-optimal represents the parameter sets of these models. The loss function \(L_p : \mathcal{F} \rightarrow \mathbb{R}\) is defined as the risk associated with the distribution \(p(x, y)\). The set \(\mathcal{F}\) is defined as
\[
\mathcal{F}_{\text{expectation}} = \{ f \in \mathcal{F} \mid L_p(f) \leq \epsilon \}
\]
Put differently, the $\epsilon$-optimal set encompasses parameter sets of the model that can achieve expected performance within the distribution $p(x, y)$. The parameter $\epsilon$ is utilized to modulate the degree of fitting, with a preference for not employing excessively large values of $\epsilon$. It is crucial to note that a model's predictions on samples from $p_S(x, y)$---whether it be the training set $DS$ or an independent validation collection---cannot serve to distinguish between different nearly optimal functions with respect to $p_S(x, y)$. This is due to the definition that, for a given expectation value $\epsilon$, the performance of any two models $f_1, f_2 \in \mathcal{F}_\epsilon$ on $p_S(x, y)$ is nearly identical. Relying solely on source data, we lack sufficient grounds to favor any specific interpretation within $\mathcal{F}_\epsilon$. Therefore, our belief system should aim to comprehensively cover $\mathcal{F}_\epsilon$, assigning non-zero significance to a variety of parameter combinations that represent different interpretative angles. This approach aligns with existing principles for reasoning under uncertainty, such as the maximum entropy principle \cite{keynes2013treatise,jaynes1957information}, advocating for evenly distributing one's beliefs across all possibilities consistent with the available information in the absence of complete knowledge.

\subsection{Training Diverse Experts }
In most practical situations, rather than meticulously selecting a neural network, the quality of data is more vital. High-quality data signifies the provision of the richest information possible to a deep learning model. In cross-domain tasks, the quality is a relative concept. If the training data can encompass as much information from the target domain as possible, in comparison to the current data domain, the model will be able to achieve an effective fit and transition to the target domain.
Specifically, when a source dataset lacks certain information, the optimal solution is to provide data containing this missing information rather than supplying a large volume of well-distributed scientific data. Although such evenly distributed data includes the information absent in the source domain, it can lead to an imbalance between samples that should be learned and those that are unnecessary for deep learning models. Consequently, models may become overwhelmed by the abundance of extraneous information, resulting in poor learning outcomes on the scant missing information.
When we want to obtain the most effective information in the target domain relative to the source domain, we do not directly analyze the target domain, but instead comprehensively assess the information already existing in the source domain.
We cover the explanations on the source domain as comprehensively as possible, 
and obtain a set of models, \(f_1, f_2, ...,f_n\) , we ensure they reach the optimal set \(F_\epsilon\) on \(p_S(x, y)\), while also ensuring that \(f_1, f_2, ...,f_n\) on \(p_S(x, y)\) avoids redundancy.
To identify these pieces of information, we introduce the concept of symmetric difference, which is the result of the union of information from the source and target domains minus their intersection. The symmetric difference in our task is defined by the equation:
\[ (f_1, f_2, ...,f_n)\,\triangle \,\,Target \,\,Domain \]
where \(\triangle\) represents calculating the symmetric difference.
In this experiment, we primarily focus on the symmetric difference found within the target domain.

Sample Evaluation Based on the Diverse Hypothesis Collection: This component utilizes the hypothesis collection to assess which samples are crucial in reducing ambiguity during the model's learning process. It entails identifying samples that can most effectively aid the model in distinguishing between these hypotheses, ensuring that the model maintains a focus on universally effective features during the transition from the source domain to the target domain, while avoiding overfitting to domain-specific salient features.
When all data participates in the training and the model optimizes to meet an expected \(\epsilon\) via the loss function, this collection of parameters of the model can be considered as the model's interpretation of the source.
In neural networks, the final stage involves passing a latent vector \( h^{\textit{now}} \)through a fully connected layer, which is precisely the best mapping of these parameters.
Ultimately, we propose a method based on the multi-perspective collection \(f_1, f_2, ...,f_n\) derived from the original dataset, utilizing this collection to assess which samples should be introduced to eliminate the current ambiguities.
This method aims to guide the model in summarizing what it has learned from various perspectives and explicitly identifies what the model needs to learn in the target domain. 

First, without introducing constraints, we train the first expert:
\begin{equation}
\small
L = \frac{1}{N} \sum_{j=1}^{N} \left( \sum_{i=1}^{M} -\left[y_{(j,i)} \log(p_{(j,i)}) + (1 - y_{(j,i)}) \log(1 - p_{(j,i)})\right] \right)
\end{equation}

where\(N\) is the total number of samples and\(M\)denotes the total number of frames within each sample.
\(y_{(j,i)}\) is the true label of the \(i^{th}\) frame in the \(j^{th}\) sample, where \(y_{(j,i)} \in \{0,1\}\).
\(p_{(j,i)}\) is the probability predicted by the model that the \(i^{th}\) frame in the \(j^{th}\) sample belongs to the positive class.\(\log\) denotes the natural logarithm.

After this, the training process, built upon the foundation of existing experts, introduces constraints with the objective of ensuring that the new interpretations sought by the neural network model in the parameter space are as distinct as possible from the existing interpretations. This is achieved by ensuring that the latent vector  $h^{\textit{now}}$ , which is processed by the neural network right before its final layer, is as dissimilar as possible from the latent vectors $h^{\textit{pre}}$ of previous experts. In other words, it aims to make the current expert's interpretation of a given input data as different as possible from those of the existing experts.
To facilitate this process, we introduce an inverse reverse distillation loss below:
\begin{equation}
L_{\textit{Dis}} = \frac{1}{n} \sum_{w=0}^{n} y_o \cdot \frac{1}{2} \max(0, \cos(h^{\textit{now}}, h^{\textit{pre}}_w) - u)^2
\end{equation}
where, \(h^{\textit{now}} \) represents a latent vector before being fed into the Fully Connected(FC) Layer layer, belonging to the current training process, while \(h^{\textit{pre}} \) represents the latent vector of a previous expert.
\, \(y_0\) \, indicates whether \(h^{\textit{now}}\) and \(h^{\textit{pre}}_w\) are dissimilar , \(\cos(h^{\textit{now}}, h^{\textit{pre}}_w)\) is the cosine similarity between \(h^{\textit{now}}\) and \(h^{\textit{pre}}_w\), \(u\) is the margin, a threshold beyond which the loss for dissimilar pairs starts increasing.

in the Given two vectors \(h^{\textit{now}}\) and \(h^{\textit{pre}}_w\), the cosine similarity between them can be calculated as:
\begin{equation}
\cos(h^{\textit{now}}, h^{\textit{pre}}_w) = \frac{\sum_{k=1}^{d} h^{\textit{now}}_k h^{\textit{pre}}_{w,k}}{\sqrt{\sum_{k=1}^{d} (h^{\textit{now}}_k)^2} \cdot \sqrt{\sum_{k=1}^{d} (h^{\textit{pre}}_{w,k})^2}}
\end{equation}

where \(d\) is the dimensionality of the vectors, \(h^{\textit{now}}_k\) is the \(k\)-th component of the \(h^{\textit{now}}\) vector, \(h^{\textit{pre}}_{w,k}\) is the \(k\)-th component of the \(h^{\textit{pre}}_w\) vector.

And the total loss function is updated as follows:
\begin{equation}
\footnotesize
L = \frac{1}{N} \sum_{j=1}^{N} \left(\sum_{i=1}^{M} \left[-y_{(j,i)} \log(p_{(j,i)}) + (1 - y_{(j,i)}) \log(1 - p_{(j,i)})\right] + L_{\text{Dis}} \right)
\end{equation}

\subsection{Selecting The Most Informative Samples}
Ultimately, we have acquired a series of well-performing, diverse experts, which represent interpretations from multiple perspectives of the original domain. Through this array of experts, we assess the informational content of the available data within the target domain, with the aim of identifying samples with the highest amount of information. Based on the information entropy, the total information quantity $I$ of a single sample, which consists of $M$ frames, can be calculated using the following formula:
\begin{equation}
I_j = \sum_{i=1}^{M} \left( -P_i \log_2(P_i) - (1 - P_i) \log_2(1 - P_i) \right)
\end{equation}
$P_i$ is the prediction result of $n$ experts, which represents the probability of the $i$-th frame being predicted as manipulated. 
\begin{equation}
P_i = \frac{m}{n}
\end{equation}
where \(P_i\) represents the probability, calculated as the number of experts ($n$) predicting the frame as manipulated divided by the total number of experts ($n$)

After calculating the information quantity, we sort the samples based on their entropy levels. For the required number of samples, we select them from the highest to lowest information quantity.

\subsection{Unsupervised Training}
In the processing unlabeled samples from the target dataset for voice activity detection, 
When two audio segments are concatenated, the distinct differences in their amplitude levels, phases, or waveform shapes often result in abrupt transitions at the juncture. These abrupt transitions introduce instantaneous high-frequency components into the signal, leading to rapid changes in energy.
Thus, our method concentrates on the absolute energy change from one frame to its predecessor and then selects the location with the highest energy change as the cutting point.
\begin{equation}
E = \sum_{i=0}^{N-1} |x[i]|^2
\end{equation}
\begin{equation}
\Delta E[i] = |E[i] - E[i - 1]|
\end{equation}

\begin{algorithm}[tb]
\caption{Samples mining with Diversity and Entropy}
\label{alg:sample_mining}
\textbf{Input}: Training data from different datasets, $h^{\text{now}}$ (the hidden-vector of the current expert), $h^{\text{pre}}$ (the hidden-vector of the previous expert) \\
\textbf{Parameter}: Loss function $L$, parameter $u$ is set to (0,1), $n$ (the number of existing experts), $z$ (the number introduced from target dataset), $R$ (the number of all samples in target dataset) \\
\textbf{Output}: Updated dataset with labeled target samples
\begin{algorithmic}[1] 
\FOR{every sample $x_{(j,i)}$ in source dataset}
    \FOR{every frame $i$}
        \STATE $L(j, i) = y_{(j,i)}(\log(p(j,i))) + (1-y_{(j,i)})\log(1 - p_{(j,i)})$
    \ENDFOR
    \IF{$n = 0$}
        \STATE $L = \sum_{j}\sum_{i} L(j,i)$
    \ELSE
        \FOR{every expert $w$}
            \small
            \STATE $\cos(h^{\textit{now}}(j), h^{\textit{pre}}_{w}(j)) = \frac{\sum_{k=1}^{d} h^{\textit{now}}_k h^{\textit{pre}}_{w,k}}{\sqrt{\sum_{k=1}^{d} (h^{\textit{now}}_k)^2} \cdot \sqrt{\sum_{k=1}^{d} (h^{\textit{pre}}_{w,k})^2}}$
            \STATE $L_{\text{Dis}}(j) = \frac{1}{2n} \max(0, \cos(h^{\text{now}}, h^{\text{pre}}_w) - u)^2$
        \ENDFOR
        \STATE $L = \sum_{j}\sum_{i} L(j,i) + \sum_{j} L_{\text{Dis}}(j)$
    \ENDIF
\ENDFOR
\FOR{each sample $x_{(j)}$}
    \FOR{each frame $f_i$}
        \STATE $P_i = \frac{m}{n}$ \COMMENT{where $m$ is the number of experts predicting the frame as manipulated}
        \STATE $I_j \gets -\sum_{i}[P_i \log_2(P_i) + (1 - P_i) \log_2(1 - P_i)]$ 
    \ENDFOR
\ENDFOR
\STATE $H = \text{Sort}(I_1, I_2, I_3, \ldots, I_R)$ \COMMENT{$H[0] \geq H[1] \geq \ldots \geq H[R-1]$}
\STATE $H = H[:z]$

\FOR{every sample $x{(i)} \in H$}
    \FOR{every frame $f_i$}
        \STATE $E \gets \sum_{i=0}^{N-1} |x[i]|^2$
        \STATE $\Delta E[i] \gets |E[i] - E[i - 1]|$
    \ENDFOR
\ENDFOR
\STATE Randomly select two groups of start and end 
\STATE Exchange their places and form labels based on the exchanged regions
\STATE Introduce these labeled target samples to source dataset
\end{algorithmic}
\end{algorithm}

where \(x[i]\) represents the amplitude of the \(i\)th frame, and \(N\) is the total number of frames in the samples. This step helps us assess the signal's strength, providing a basis for subsequent analysis.

Next, we calculate the zero crossing rate which is an important metric for assessing changes in signal frequency.
By calculating the absolute change in energy \(\Delta E[i]\) between adjacent frames and zero crossing rate, we can more accurately determine the start and end points of voice activity. During silent segments, the energy change is minimal; however, at the onset or conclusion of speech, the energy change is significantly larger. Therefore, by randomly selecting points where the energy change exceeds a threshold as the cutting points for active voice segments, and then randomly selecting the start and end points of two segments from these threshold-exceeding points, followed by swapping these two segments, we can label the entire audio. Areas without exchanges are labeled as 1, while exchanged areas are labeled as 0.

\section{experiment}
\subsection{Baseline and Dataset}
\subsubsection{Linear Frequency Cepstral Coefficients}
 LFCC is an acoustic feature extraction technique used in speech processing that effectively represents the features of speech signals.
\subsubsection{Light Convolution Neural Network}
LCNN consists of convolutions, max-feature-map (MFM) modules, a pooling layer, and a linear layer. The MFM is a module that only retains the maximum value of the two channels to reduce CNN architecture. A convolution and an MFM module are combined to form a group. Several groups are arranged consecutively to encode the input feature, and then the encoded feature is pooled and sent to the linear layer to obtain scores. More details about LCNN can be found in \cite{lavrentyeva2017audio}.
\subsubsection{PAF dataset}
In our experiments, we use Half-truth Audio Detection (HAD) dataset \cite{yi2021half} and ADD2023track2 dataset.The HAD is generated based on the AISHELL-3 corpus \cite{shi2020aishell} which consists of 88035 utterances about 85 hours from 218 native Chinese mandarin speakers. The authors use fake clips generated by TTS and vocoder to replace named entities or attitudinal words to generate fake audios. 
The HAD dataset includes 50000 audios in the training set, 17823 audios in the validation set, and 50092 audios in the test set.

For the ADD2023track2 dataset, by providing a mix of genuine and manipulated audio segments,its test set contains genuine and synthetic voices from six different datasets, including Ai-shell3 Ai-shell1 etc. 
The ADD test set includes 53092 audios.The training and validation sets of it are adopted from those of HAD.

\subsection{Training Detail and System Setting}
The signals are divided into frames with a frame length of 10ms to extract LFCC features, and the overlap between adjacent frames is 10ms. To control variables, all subsequent experimental comparisons will be based on LFCC as the feature input. 
The Adam optimizer, with a learning rate of 0.0001 is used with a batch size of 16 utterances. Note that a segment with 750 frames will be intercepted from the feature if it has more frames than 750 while smaller ones are zero-padded to match the size of the longest utterance in the batch.

During the training process, the parameter \(u\) of the reverse distillation loss, which represents the interval between the ten experts, was set to a value between 0 and 1. And the parameter\(y_0\) was set to 1. And the parameter\(n\) which is the number of experts was set to 10.
Instead of relying on the Equal Error Rate (EER) for evaluation, Track 2 incorporates multiple metrics to calculate the final score of a submission. These metrics include sentence accuracy ($A$) and segment $F1$ score, which is calculated by recall and precision. More details can be found in\cite{yi2023add}.
All of the experiment are training on HAD training set but testing on ADD2023 test set,Using LFCC as the base model, the parameter \(u\) of \(L_{\textit{Dis}}\) is set to 0.75.

\subsection{Fails on Cross-domain Dataset}

\begin{table}[t]
\centering
\caption{The F1scores (in \%) for the SDF and other models, which are trained using the HAD training set, and test on the HAD training set, HAD validation set, and HAD test set.}
\begin{tabular}{@{}lcccccc@{}}
\toprule
& \multicolumn{2}{c}{HAD dev} & \multicolumn{2}{c}{HAD test} & \multicolumn{2}{c}{ADD test} \\
\cmidrule(r){2-3} \cmidrule(r){4-5} \cmidrule(r){6-7}
& Avg. & Best & Avg. & Best & Avg. & Best \\
\midrule
RSDM & \textbf{98.26} & \textbf{99.17} & \textbf{96.01} & \textbf{97.76} & 04.74 & 09.19 \\
ADD2023baseline & 93.15 & 94.83 & 92.03 & 93.71 & 08.38 & 12.13 \\
\textbf{SDE (ours)} & 96.00 & 96.25 & 87.71 & 88.81 & \textbf{43.84} & \textbf{47.24} \\
\bottomrule
\end{tabular}

\label{your-label-here}
\end{table}

\begin{table}[t]
\centering
\caption{Ablation experiment results for the SDF model by introducing 5000 samples from target dataset, trained using the HAD training dataset and tested on the ADD datatest set.}
\begin{tabular}{lccccc}
\hline
Model & F1score(\%) & Recall(\%) & Precision(\%)  \\
\hline
\hspace{5mm} SDF & 43.84 & 49.12 & 39.58  \\
\hspace{1mm}w/o - diverse experts & 23.59 & 29.31 & 19.73  \\
\hspace{1mm}w/o - evaluating entropy & 11.23 & 9.50 & 13.73  \\
\hspace{1mm}w/o - labeled samples & 15.53 & 18.04 & 13.63  \\

\hline
\end{tabular}

\label{table:ablation_study}
\end{table}

We have evaluated our method with two currently available models for locating manipulated regions on in-domain datasets, which have already published their code. RDASDM refers to “a Robust Deep Audio  Splicing Detection Method via Singularity Detection Feature”\cite{9746596}, and The ADD2023baseline originates from \cite{yi2023add}. All models utilize the LCNN architecture as their foundation. 
We set the shuffle parameter of the dataloader function to true, conducted two experiments on each of the three models, and reported both the average and the best values.
However, the RDASDM model differs in that it incorporates wavelet features and Long Short-Term Memory model, making its structure more deep and complex.
\begin{table*}[ht]
\centering
\caption{The average and best F1score (\%) of various mining methods at different data sizes, trained on the HAD training set and tested on the ADD 2023 test set.}
\label{tab:results-data-size}
\begin{tabular}{@{}l@{\hspace{20pt}}c@{\hspace{10pt}}c@{\hspace{20pt}}c@{\hspace{10pt}}c@{\hspace{20pt}}c@{\hspace{10pt}}c@{\hspace{20pt}}c@{\hspace{10pt}}c@{\hspace{20pt}}c@{\hspace{10pt}}c@{}}
\toprule
& \multicolumn{2}{c}{2500} & \multicolumn{2}{c}{5000} & \multicolumn{2}{c}{7500} & \multicolumn{2}{c}{10000} & \multicolumn{2}{c}{20000} \\
\cmidrule(r){2-3} \cmidrule(r){4-5} \cmidrule(r){6-7} \cmidrule(r){8-9} \cmidrule(r){10-11}
Method & Avg. & Best & Avg. & Best & Avg. & Best & Avg. & Best & Avg. & Best \\
\midrule
Undersampling & 11.07 & 13.02 & 14.34 & 16.91 & 18.07 & 19.84 & 18.64 & 21.82 & 40.68 & 41.04 \\
Oversampling & 14.07 & 14.69 & 16.14 & 16.67 & 19.07 & 20.78 & 23.64 & 24.40 & 42.68 & 44.98 \\
Negative Mining & 07.54 & 7.65 & 08.41 & 8.63 & 10.57 & 11.07 & 14.93 & 15.11 & 36.82 & 37.45 \\
Multi-Cluster & 23.39 & 24.17 & 24.74 & 26.44 & 32.02 & 33.36 & 32.16 & 33.17 & 49.48 & 55.25 \\
Random & 10.44 & 18.01 & 13.11 & 22.29 & 18.56 & 27.25 & 23.54 & 33.79 & 37.22 & 50.44 \\
\textbf{SDE (ours)} & \textbf{28.33} & \textbf{29.21} & \textbf{43.84} & \textbf{43.98} & \textbf{51.64} & \textbf{51.73} & \textbf{56.99} & \textbf{57.77} & \textbf{67.68} & \textbf{69.92} \\
\bottomrule
\end{tabular}
\end{table*}

From Table 1, it is evident that when faced with cross-domain datasets two of the models failed. Additionally, our model with a simple structure shows slightly poorer fitting on the source dataset than the other two models. We hypothesize that the probable reason for this is that our network architecture is not as deep and our neural network is shorter than those of the other two models.

It's noteworthy that the F1score is an indicator of model performance and represents the model’s fit. In in-domain tasks, a higher F1score generally indicates better performance. However, for cross-domain tasks, a higher performance in the source domain could potentially lead to failures in the target domain.

\begin{figure*}[htbp]
  \scriptsize
  \centering
  \caption{The left side shows the average entropy size for all samples of the target dataset with different parameter settings of u, and the right side shows the F1score results for the corresponding u parameter settings when introducing 5000 samples from target dataset.} 
  \begin{minipage}[b]{0.49\linewidth}
    \includegraphics[width=\linewidth]{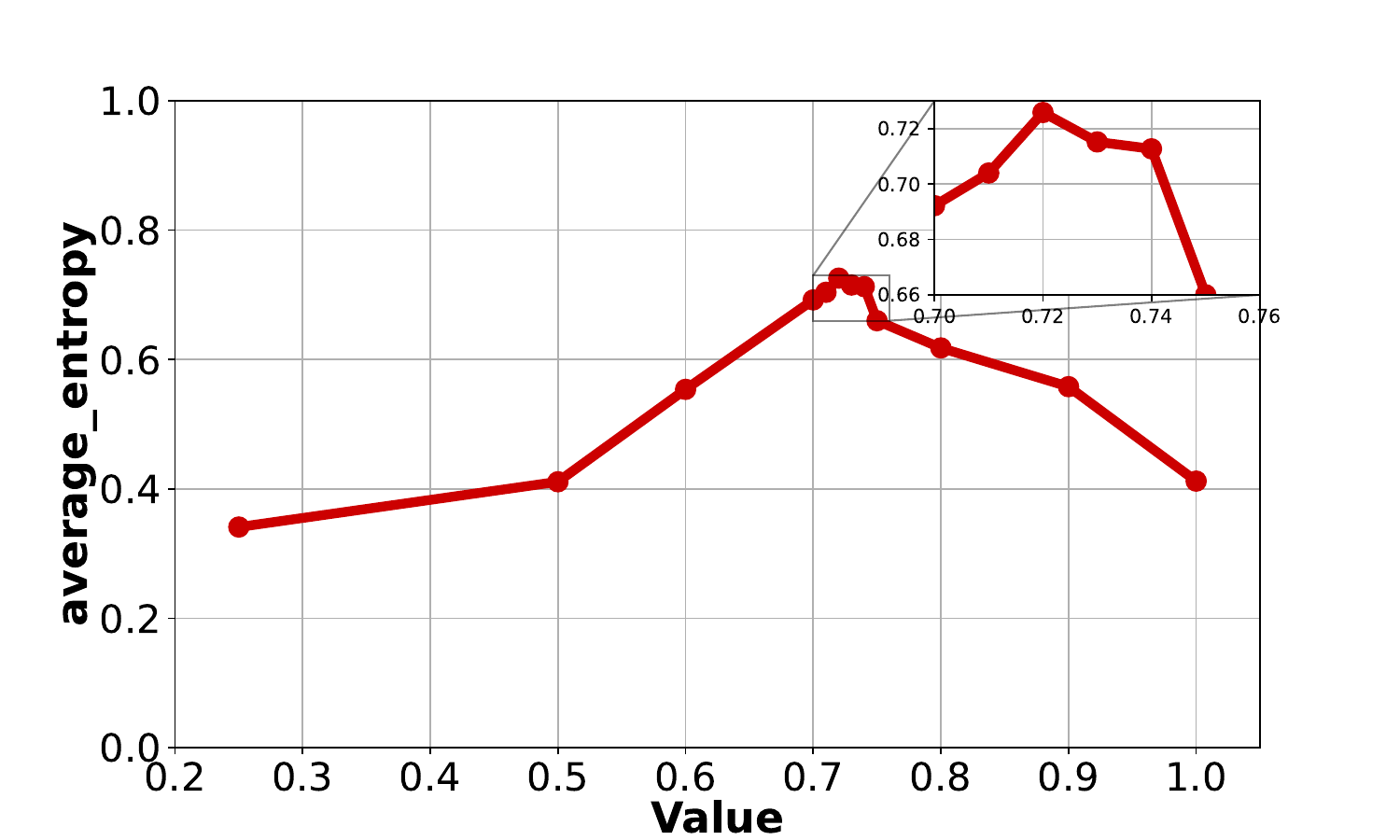}
    \label{fig:average_entropy} 
  \end{minipage}\hfill
  \begin{minipage}[b]{0.49\linewidth}
    \includegraphics[width=\linewidth]{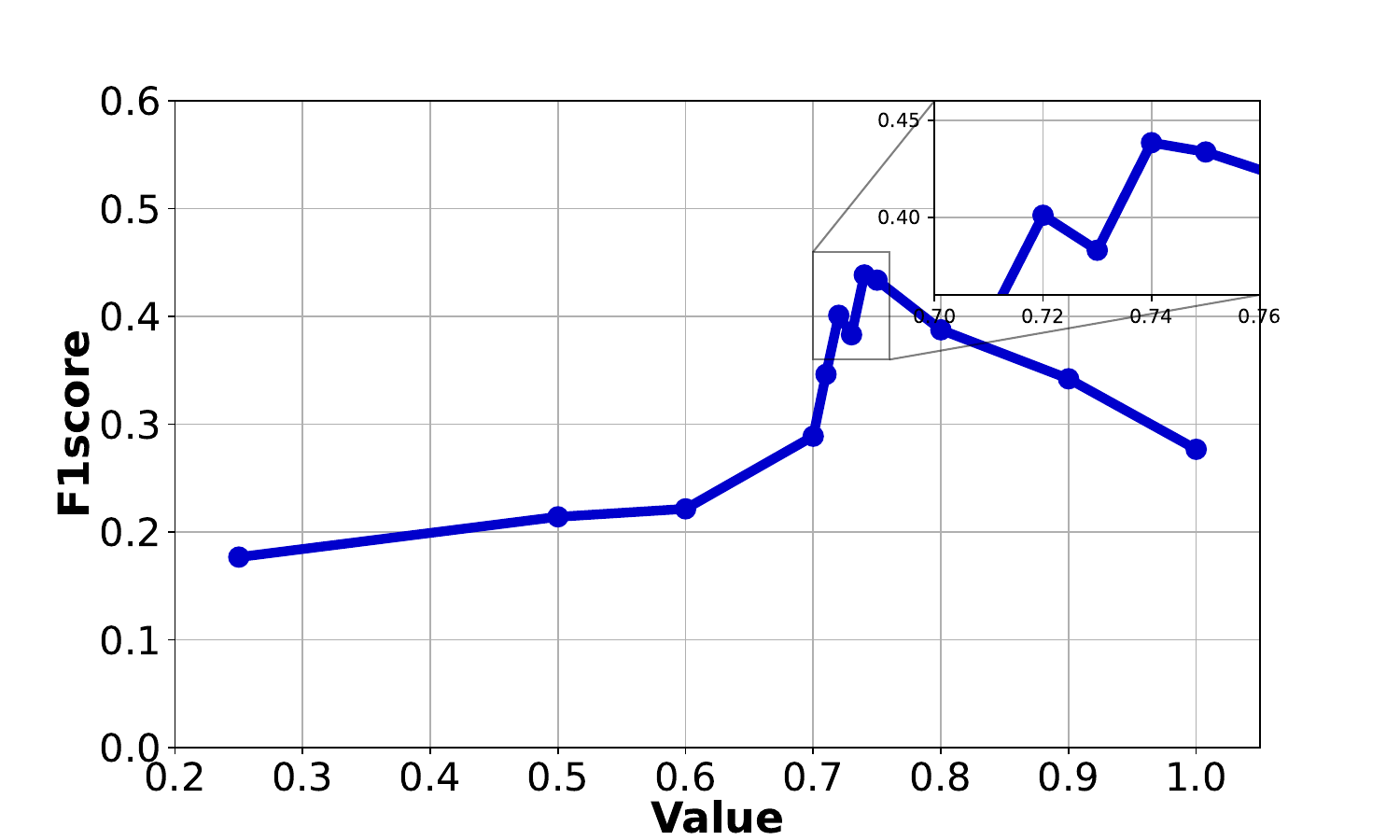}
    \label{fig:f1_score} 
  \end{minipage}
  
  \label{fig:combined_figure} 
\end{figure*}

\subsection{Ablation Experiment}
In the ablation experiment, we evaluated the contribution of each method within the SDF model. The "without diverse experts" configuration was set to randomly shuffle the training data without adding any other constraints. The "without evaluating entropy" configuration was set to randomly introduce samples from the target dataset. The "without labeled samples" configuration was set to not detect voice activity or swap segments, and to train using randomly generated labels for each introduced sample. The results are shown in Table 2.
When the diverse experts were removed from our method, performance dropped by 46\%. The main reason for this decline is the absence of experts to assist in sample selection, which hinders the effective selection of samples. However, due to the shuffling of the dataset, models optimized from different starting points may still retain some capability. In fact, most of the samples we selected were those with a prediction probability \(P = \frac{1}{10}\). Which means that at this point, most of the experts are homogeneous.
In the scenario without evaluating entropy, we were only able to select samples randomly. Consequently, the final experimental results showed a significant decline in performance, amounting to a 74\% drop.
when random labels were used, performance declined by 64\%. We speculate that this could be because random labeling may lead to the misclassification of boundary frames that the model should learn, thereby causing a decrease in model performance.
\subsection{Comparison of Methods for Mining Samples}
In the study, we introduce a series of experiments to evaluate the effectiveness of SDE. We take LFCC as input, use LCNN as the base model, train on HAD training set but test on ADD2023 test set and employ the F1score as the evaluation metric.
We first experimented with the performance of various sample mining methods when a certain number of target dataset samples were introduced. Among them, Multi-Cluster employs the K-means clustering method to analyze the existing samples in the target dataset. It selects samples based on their proximity to the cluster centers, with the number of clusters set at 6 in this case. Undersampling, using the same analysis method as Multi-Cluster, randomly selects samples from sparsely distributed areas. Conversely, Oversampling also based on the same analysis as Multi-Cluster, randomly selects samples from densely populated areas. Negative Mining, on the other hand, intentionally selects samples that the model consistently misclassifies with the same outcome, representing the samples with the lowest information entropy. 
It is noticed that, in order to make the experimental results more objective, we conducted two or more experiments for each method and recorded the outcomes. Specifically, we performed five experiments using the random method. For our method, SDE, we conducted two reverse distillation processes. Since we set up shuffling in the dataloader, we can consider the experts generated in the two iterations to be different. The other methods were tested three times each.

Under-sampling mitigates class imbalance by reducing the number of examples from the majority class, which simplifies the model training process and potentially diminishes the risk of overfitting. We have employed under-sampling specifically on the majority class within the target dataset. Similarly, multi-cluster sampling ensures diversity in the sampling process and enhances the model's adaptability to different data distributions by partitioning the dataset into several clusters and sampling from each cluster. We hold great expectations for these two methods.
The results, surprisingly, are clearly evident in Table 2: neither method managed to select high-quality samples, as seen from the F1score outcomes. The results from random selection met expectations, showing that the F1score increase is nearly proportional to the number of samples added.

Furthermore, as indicators based on information entropy, two completely opposite methods, Negative Mining and our method SDE, demonstrate the effectiveness of entropy-based sample selection strategies. Firstly, when a limited number of samples are introduced, the outcomes of our method surpass those of Negative Mining. Additionally, the data in Table 2 clearly shows that for our method, the slope of the performance improvement line for the top 10\% of samples is steeper than that for the top 20\% of samples. This suggests that the samples in the top 10\%, which have higher information entropy, are of higher quality than those between 10\% and 20\%.
Our method significantly improves model performance by addressing the information disparity between the source and target domains. By selectively choosing samples that offer more information, we enhance the source domain model's comprehension when applied to the target domain. Two key factors contribute to the success of our approach. First, we employ reverse distillation to minimize shared information among different experts. This process compels experts to concentrate on secondary distribution information within the source dataset, enabling them to interpret the dataset from diverse perspectives. Secondly, by selecting samples from the target dataset that possess maximum information entropy relative to the source dataset, we effectively reduce the model's ambiguity. Together, these strategies optimize the transfer of relevant information across domains, leading to notable improvements in model accuracy and robustness.

\subsection{Entropy and Information}
To demonstrate the correlation between information entropy and performance, by setting different values of parameter\(u\), we learned multiple sets of experts with significant viewpoint differences in the source dataset. These experts evaluated the information entropy of the existing target samples, with the average results shown in Figure 3. Subsequently, we used these samples with varying information entropy levels for evaluation, and the experimental results are presented in Figure 3.
The parameter \(u\) serves as a constraint on the minimum distance between latent vectors, essentially acting as a distance limitation among experts. A sufficiently large and appropriate \(u\) value  ensures that the experts interpret from multiple distinct perspectives within the source dataset, rather than from similar angles.
Furthermore, more importantly, as the average information entropy increases, the F1score also rises, indicating that the amount of information entropy in the introduced samples directly impacts the model's performance. This relationship underscores the significance of selecting high-entropy samples to enhance the model's ability to generalize and perform effectively.
We subsequently evaluated the diversity among the ten experts trained with the \(L_Dis\) parameter 
\(u\) set to 0.75. This evaluation was conducted by calculating the dissimilarity in their predictions for the samples.
The results shown in Figure 4 indicate that the initial training stage, where the divergence among experts was the greatest, resulted in the best performance. The clear initial differentiation among experts enables more effective exploration of diverse strategies and insights, leading to improved overall model performance. As the number of experts increases, the parameter space becomes more complex with multiple divergence points. This complexity can lead to potential conflicts in optimization directions. Consequently, experts trained later might not diverge sufficiently from the perspectives of the original experts, leading to homogenization. 
\begin{figure}[t]
\scriptsize
\centering
\caption{The dissimilarity of predictions among experts}
\includegraphics[width=\linewidth]{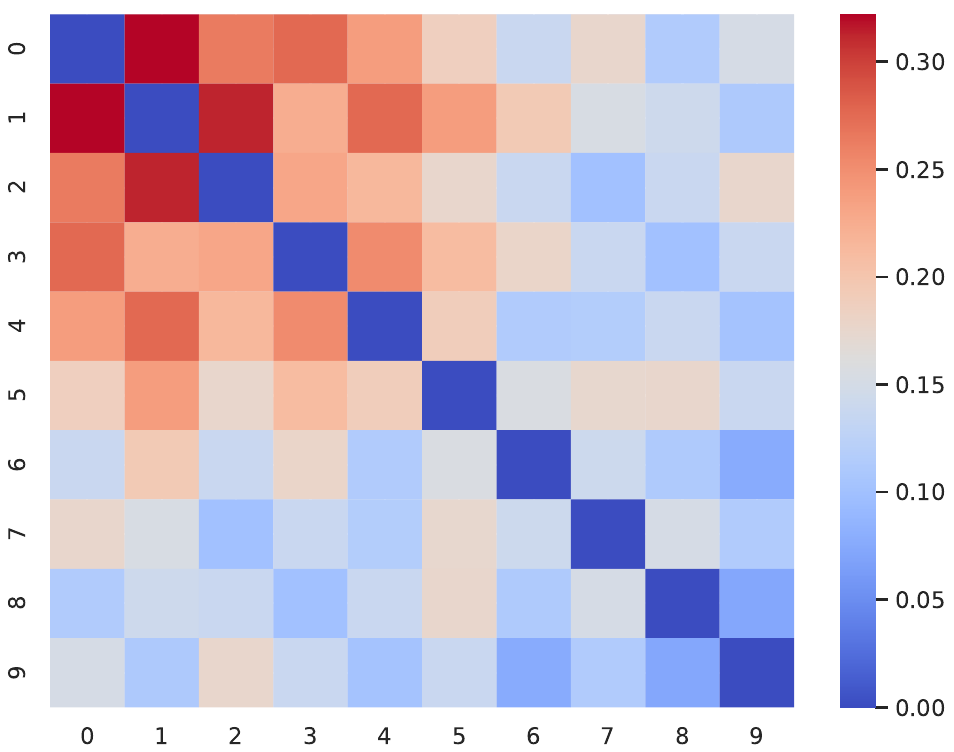} 
\label{fig:bigfigure}
\end{figure}

\section{Conclusion}
In this work, we propose an unsupervised domain adaptation method called SDE to overcome out-of-distribution challenges in the task of partially fake audio. This method introduces the most informative samples by calculating their entropy relative to the original domain. Additionally, we have designed a training method for unlabeled samples for the task of locating manipulation regions in partially fake audio. The specific operation involves selecting segmentation points based on the rate of voice activity changes to perform segment swapping and label generation. 
The generated labels and the swapped segment samples are then used for training in the source domain.Our results indicate that our method outperforms five other sample selection techniques in domain adaptation for the specific task of identifying manipulation regions in partially fake audio. In this context, by incorporating the top 10\% of samples from the target domain, we achieved an F1 score of 43.84\%. Compared to the performance of the second-best method, this score represents a relative enhancement of 77.2\%. 
Additionally, the outcomes of the Ablation Experiment elucidate the distinct contributions of each phase within the SDF methodology, thereby providing indirect substantiation of our method's efficacy. We propose a method based on entropy for mining the most relatively abundant information, which has a very broad range of applications. The most direct application scenario is the localization of tampered regions in images, which will be the focus of our future experiments.

\bibliography{base}

\end{document}